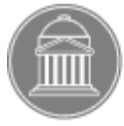
BOBBY B. LYLE
SCHOOL OF ENGINEERING

# Effective Security by Obscurity

J. Christian Smith

chris@smu.edu

12/10/2011

---

**Abstract**

"Security by obscurity" is a bromide which is frequently applied to undermine the perceived value of a certain class of techniques in security. This usage initially stemmed from applications and experience in the areas of cryptographic theory, and the open vs. closed source debate. Through the perceived absence of true security, the field of security by obscurity has not coalesced into a viable or recognizable approach for security practitioners. The ramifications of this has resulted in these techniques going underused and underappreciated by defenders, while they continue to provide value to attackers, which creates an unfortunate information asymmetry. Exploring effective methods for employing security by obscurity, it can be seen that examples are already embedded unrecognized in other viable security disciplines, such as information hiding, obfuscation, diversity, and moving target defense. In showing that obscurity measures are an achievable and desirable supplement to other security measures, it is apparent that the in-depth defense of an organization's assets can be enhanced by intentional and effective use of security by obscurity.

---

## 1. Introduction

    "Security by obscurity" (SBO) is a label of derision whose application has been most often applied in two areas, cryptography and open source. Auguste Kerckhoffs showed that any secure military system "must not require secrecy and can be stolen by the enemy without causing trouble." The early years of historical cryptography are characterized by obscurity mechanisms, and then by complexity, as opposed to true unsolvable (NP-complete) algorithms. These strategies often worked in practice for hundreds of years due to their obscurity. However, with the advent of computers in the late 20th century, cryptography based on "security by obscurity" failed in case after case where Kerckhoffs' principle had not (yet) been adopted as doctrine.

    In the open versus closed source debate, the source of the "security by obscurity" aspersion refers to the programming practice of hiding secrets in source code, releasing only the executable code, and the belief that code secrecy in general can make a system more secure [24]. Although hiding source code does provide some trivial protection from attack if secrets are stored in it, the underlying assumption, that binary encoding secures either the source which created it, or the secrets inside it, turns out to be proven false [46]. According to David Wheeler, "why would an attacker need a blueprint, when they can get an exact copy of the building to explore and search for vulnerabilities?" [8]. The lesson is rather to provide true security for your secrets, for example by cryptographically protecting them in the code, or better still, not embedding them at all.

    Considering these examples in more detail, an interesting and contradictory idea can be extracted. First, despite the weak mechanisms of historical cryptography, these systems were unbroken for hundreds of years! Modern cryptographic mistakes often take years to uncover. In the open source realm, advocates deride hiding information in

source, yet in admitting that releasing source code makes it easier to extract secrets, they also admit that hiding code makes it harder. Is there value in the "trivial protection" that obscurity affords after all?

It has been said that opening the source gives an unfair advantage to the attacker: they need to find but one vulnerability to successfully attack the system, whereas the defender needs to patch all vulnerabilities to protect himself completely [19]. From the cryptography example comes the design principle that the security of a system, should not depend on its secrecy, on the grounds that the system is shared by many other people and therefore will become public in due time. Doesn't it follow that secrets should be kept and not released? There is a gap in these arguments which is easy to overlook: until such a time as its defeat, a secret system remains secure. A value from obscurity can be obtained if there is a method to intentionally manipulate and leverage this period of security by obscurity, for example, by stretching out the time required to defeat the system.

Delving further into the lessons of Kerckhoffs, the open source debate, and a history on security by obscurity techniques, the goal of this paper is to show that security by obscurity has value in information security practice. Rather than supply formal proofs, this will be done by highlighting other work, examples that SBO thought and practice is here to stay. SBO concepts ignored by security practitioners continue to provide value to attackers, which creates an unfortunate information asymmetry. Other techniques which are already entrenched in prior security practice, go unrecognized. Some recent techniques, which few would label security by obscurity, actually do show the value of these principles, and form the cutting edge of research today.

Obscurity measures which have value may be considered useful as part of a defense in depth [3] approach, in which an organization always assumes that each individual measure may be circumvented but each obstacle along the way still strengthens the security of the whole. Exploring effective methods for employing security through obscurity, by intentional design, can be a fruitful endeavor when combined with other security measures, as part of an organization's defense-in-depth strategy.

## 2. Kerckhoffs' Doctrine

The debate about open versus closed systems was kicked off in the nineteenth century. In 1883, Auguste Kerckhoffs proposed that (translated from the French):

> The security of any cryptographic system does not rest in its secrecy, it must be able to fall into the enemy's hands without inconvenience. [...] Assume that the method used to encipher data is known to the opponent, and that true security lies in the choice and secrecy of the key. [22]

This has come to be known as Kerckhoffs' doctrine

Ironically, it has been pointed out, hiding keys is of course security by obscurity [24], but this assumption of no-secrecy in the cipher system is instead taken as a vote against obscurity as a protective measure. Kerckhoffs' doctrine is used as a touchstone in the mantra against security by obscurity. This has ultimately extended into the debate about whether access to the code of not just a cryptographic algorithm, but an entire software product, can and should be kept hidden as a security measure. Is it of more help to the defense, as the community can fix weaknesses, or to attackers, because they can develop exploits for them more easily? Kerckhoffs' idea is widely accepted among all participants in the debates. Security inventors who rely on secrecy have tended to over-rely on it, making mistakes in security implementation which are eventually discovered, often to disastrous effect. However, open source advocates extend Kerckhoffs' doctrine to support the idea that open source is good security, as it better fulfills the principle that a protection mechanism must not depend on attacker ignorance. However, a more conservative interpretation would be that it shows that obscurity alone is insufficient to provide complete security, and of course, there's no such thing as complete security!

A classic modern example of the failure of closed cryptographic systems was Content Scramble System (CSS), the digital rights management (copy protection) and encryption system used in DVD-Video discs. Released in 1996, CSS was developed by the DVD Forum, without being an open standard. By 1999, the cryptography employed was broken by a Norwegian software developer: "If the cipher was intended to get security by remaining secret, this is yet another testament to the fact that security through obscurity is an unworkable principle" [41].

They should have known better: the decision go with a secret design was in violation of Kerckhoffs' Principle and resulted in a bad implementation. In CSS, the weak choice of a short key length was a well-known mistake, yet it took years to defeat, despite a lot of motivated individuals against it. It takes time to break cryptography, and the only thing which kept the system secure for three years was the obscurity of the method used. It's hard to argue with the fact that obscurity provided three years of security value. In this case, however, the marginal cost to provide more than three years of security, stronger cryptography, was negligible, and the requirement for long-term security was extremely high. In other cases, if the obscurity measures are cheap relative to the duration of protection, a entirely different conclusion could be reached, that it was good-enough security (and cheap!). This idea will come up again.

While many participants take a position of orthodoxy regarding Kerckhoffs' doctrine, treatments that attempt to delve deeper into whether obscurity *never* helps or whether open peer review is *always* superior (or even usually superior) tend to come up ambivalent [4], [23], [36]. For example, comes this: "building a fault-tolerant silk purse out of less robust sow's ears is indeed possible in some cases" [27]. The lack of a simple and compelling proof of Kerckhoffs' in all cases invites further discussion of the merits of security by obscurity.

## 3. Security by Obscurity and the Open Source Debate

The open vs. closed source debate is virtually a religious issue in computer science. One of the gods of the members of the pantheon, Richard Stallman (of GNU), espoused the seminal belief that not being able to change or share programs is fundamentally unethical [40]. What started as a "stark moral choice," over time turned into an activist mission for free and open source software, yet that core ethical dilemma has to do with neither business models nor source code.

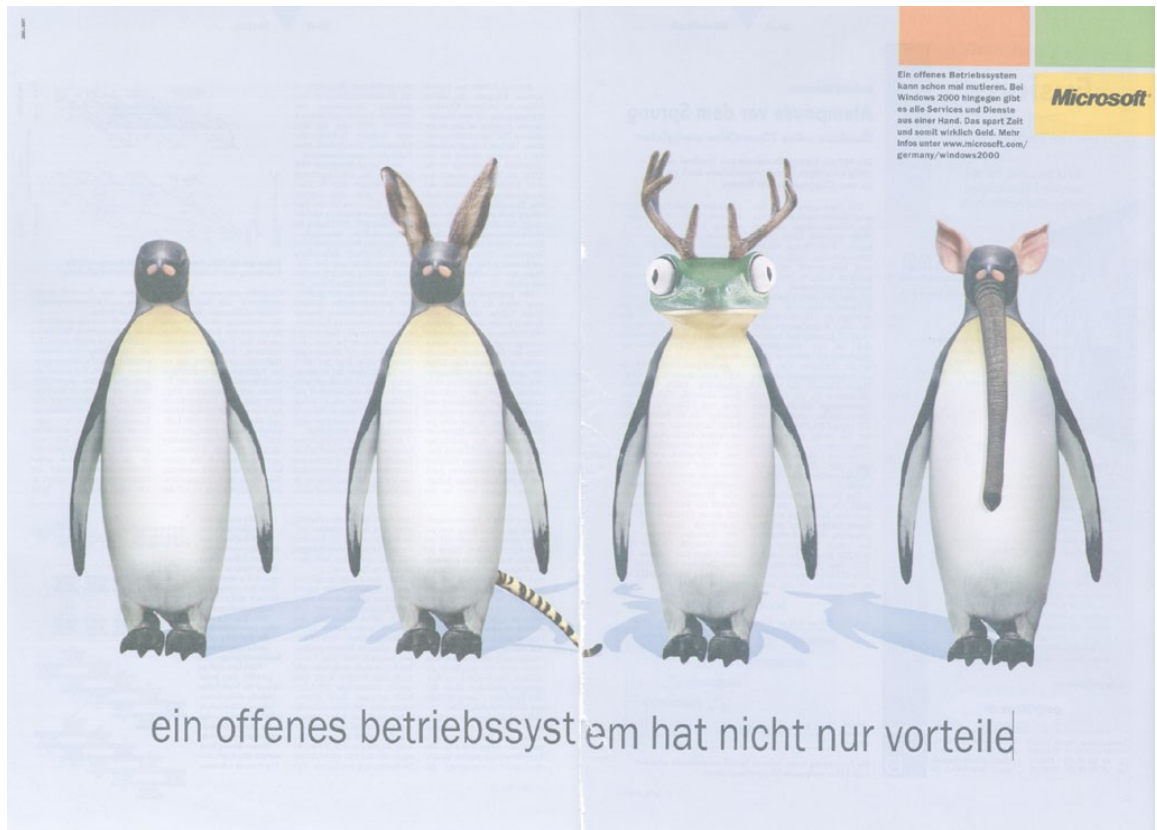

Figure 1. Microsoft's attack on openness and Linux. *c't*, October 2000.

The *Cathedral and the Bazaar* [32] is an essential treatise from another legendary forefather of the debate, Eric Raymond. The cathedral model of software development, used by everyone, from Stallman's own Free Software Foundation (FSF), to proprietary companies (e.g. UNIX), is based on coordination, strategy, planning, and goal-setting in software development. In the bazaar model, exemplified by the development of Linux, open source comes from the dis-coordinated yet efficiently delegated distribution of labor, yet there is evidence it can result in technically superior products.

Humorously characterizing Stallman and Raymond as the beneficent gods of Open Source, invites labeling Microsoft as the wicked trickster god, Loki. Microsoft has attacked open source multiple times in the past, either through scholarly venues [23], or in the public forum. In one humorous example criticizing the unreliability of mutation, they show that "an open operating system has not only advantages" (Figure 1).

Although humorous in this picture, attacks like this highlight the polarizing nature of the debate, and far more vicious ones abound in the literature -- remember "Microsoft, the Evil Empire?" Microsoft own published research on the source code release issue spread fear:

> The "dark side" of access to source code is enhanced opportunity for malicious parties to develop attacks based on detailed knowledge of flaws.... Access to source code can provide the leverage to build a subtle and sophisticated attack against a widely deployed system. [...] Disclosing the source code of a product may expose it to subtle and malicious attacks that greatly jeopardize its user base. [23]

References for these assertions were absent.

Some early opponents of the many-eyes ("given enough eyeballs, all bugs are shallow" [32]) open source benefit on security, believed that the invalidity of this alone was enough to discredit open source benefits altogether [36]. In this view, bugs in code are not the dominant avenue of attack, and they are solvable by secure programming (type correctness, exception handling, etc...) and not cause for embracing open source. However, history subsequently showed that open source products were far more secure than their commercial counterparts [26], and more widely adopted after all in key areas needing security, notably internet services [47].

More than a religious issue, the source code disclosure vs. information hiding debate has also become a political issue. From media coverage of US attempts to monitor electronic communications, and concerns over the lack of transparency in efforts to combat terrorism, has come the perception that commercial software may be forced to incorporate secret back doors in their software to permit and facilitate monitoring. Access to source code, for commercial products, is one proposed way to prevent this.

At the risk of navigating into the icy waters of the open source debate, despite the proven efficacy many-eyeballs, there is an element of risk, and an element of faith-based security and trust associated with publicly releasing code. Taking the most robust example, cryptography is an area in which sharing code widely is essential and proven to incur positive results through having other cryptography researchers weigh in on its security. In publicly releasing a new algorithm, a researcher trusts many things, examples such as these:

1. That the community will successfully vet the security of his algorithm (they are not obligated)
2. That a malicious individual, group, or nation, won't keep knowledge of weaknesses secret for future profit (they might)
3. That another individual won't profit off, or weaponize his work (they might)

In the cost-benefit analysis, the researcher concludes the likelihood of these negative events occurring is small, and the benefit of sharing is enormous: it's perhaps the only way for an algorithm to become trusted. On the other hand, a proprietary, secret algorithm, kept obscure, doesn't gain these advantages, and as a result case after case of cryptography security by obscurity failures exist, as we have seen. Such cryptography cases are the smoking gun [6] in the death of security by obscurity as a viable technique. However, the logic behind these arguments is tightly tied to aspects unique to the field of cryptography. The risk remains extremely small in this case because the community is focused on and motivated to provide the vetting, for example by recognition or publication [23], and few or none of the individuals outside the tiny cadre of collegial researchers have the requisite knowledge required to find weaknesses which the rest of the group has missed. These assumptions don't appear to hold equally true, at least they are not guaranteed, for source code disclosure. Yet many authors negatively equate source code secrecy with security by obscurity, either indirectly or directly: "The belief that code secrecy can make a system more secure is commonly known as security by obscurity" [24].

As mentioned, proving that code promiscuity can make a system more secure has already been done [26]. In reality, though, finding security holes in software is difficult, particularly for those uninvolved in the development of the system [46]. This distinguishes a commercial product from an open source one in which "many eyes" not only may review code, they are also active participants in the development model of the bazaar, which works in favor of their efforts. And, of course, the source code does not reveal all problems in object code, an important consideration where it is used to validate trust in accompanying a binary distribution.

In light of this, one opinion on the debate seems quite poignant. When an open source company publishes source code, they are allowing the community to improve the code, which offsets the risk of attackers being more or less (depending on your opinion) aided in their efforts by having access to the code. However, in most cases when a proprietary code company publishes source code to something you use, they immediately weaken your security, unless you have the capability of rapidly finding, fixing, and rebuilding the software faster than an attacker can leverage it. Plus, a proprietary license won't allow you to fix it [21]. This distinction makes it clear that a commercial proprietary company cannot in fairness be asked to release its code, but, rather, to change its licensing, developmental, and business models.

It's also been noted that open source is quite different from open information, and it is important to keep in sight when the "many eyes" benefits are a detriment, as in the case of protecting an organization's proprietary information.

Stuttard says: "An organization will not make itself or anyone else more secure by releasing information about its own infrastructure, technologies or configuration" [42].

Some argue that regardless of release of code, using the open-source model internally leads to better security-aware development, "software developed knowing the source will be open is more secure" (John Pescatore) [25]. This is thought because more people study the code, because the open development community codes better, or is more aware of security problems, or just because the transparent development model lends itself to better security through openness and public accountability. These arguments assume the poor quality of commercial-software development. Open source proponents point out that companies license the right to use their products under the condition that they are not liable for damages, which de-motivates them and ensures their developmental practices don't favor security over productivity problems. This effect perhaps is balanced by the counterargument that true accountability is financial ("money talks") and companies are more easily incentivized to deliver secure products if the marketplace calls for it.

Ultimately, this debate is intrinsically tied up with security by obscurity, as security experts have abused the term in discrediting the idea that keeping source code secret makes it harder for malicious agents to develop malicious code, on the grounds that it is still possible to use object code, and secrets always come out in the end. The opposing argument, that releasing code makes it easier to find vulnerabilities, carries with it a large degree of intuitive weight, particularly when one considers code-discovered vulnerabilities, and the efficacy of code pattern-matching compared to static attacks on binaries. It is logical to argue that releasing code particularly enables the automation of seek-and-destroy style code-based attacks. Research has underlined this, showing the viability of automated tools to find vulnerabilities in source code [37].

Of course, automated of source code review, works for defense as well, and there are many free open source tools available. Ironically both sides of the debate are actually in agreement, recognizing that disclosing the source makes it easier to find security holes in code, either to fix or to attack! Many analysts have pointed to open source as an opportunity to raise the security bar arbitrarily high as users can optionally integrate in multi-source security-enhancing technologies, beyond what a single closed-source vendor can offer, as well as perform their own code validation [10].

There is often a difference in perspective represented between the two sides of this debate which deserves mention. Open source advocates call frequently for *external* code to be released open source, so that an organization can audit and extend code, trust and/or validate that it was developed securely. This is a receive-open perspective. On the other hand, closed source advocates display the perspective that *internal* code that an organization creates or uses which is sensitive should not be readable by third parties. This is a send-closed perspective.

This property has been highlighted as a serious problem and an example of the divergence of the two sides:

> Open source is, by definition, designed to be world-readable, so no secrets are embedded in it. Closed source, on the other hand, comes with the temptation to embed secret data, and to achieve any degree of security by means of clever tricks, and not through proven algorithms. [13]

The assumptions behind such a statement reveal the biases of those opposed to closed source with near-irrational fervor. Not only is it illogical to declare that there are never secrets embedded in open source software (particularly if they are encrypted!), but assessing the validity of the temptation to which the author alludes is a job for psychologists and not computer scientists.

Instead, we can more constructively choose to see the arguments on both sides, rather than divergent, as mutually compatible. Surprisingly, the two ideas, combined, conform to the same mandatory access control model, the Simple Security Property of Bell-LaPadula [44]. In the open sourcer's case they require that an object of lower security trust level (public source code) should be readable from a higher level (defender), whereas in the closed sourcer's example, an individual with lower security trust level (attacker) may not read information at a higher level (defender's code).

An analysis of the arguments perhaps reveals that both perspectives are right. Security by inobscurity (many eyes), and security by obscurity are not entirely incompatible. While secrets should not be hidden in code, there is some value for some period of time to keeping code secret, a value which has to be weighed case-by-case against the disadvantages to themselves of an organizations' closed source developmental model, and the disadvantages to the community of keeping the source closed. In the end, there are other far more compelling methods of leveraging the security by obscurity principle, and hiding code is among the least useful.

## 4. The War over Disclosure

Any discussion of open source and vulnerabilities has to touch upon full-disclosure. The Bugtraq open list flourished from 2001-2005 and was a source of continual controversy. While security researchers (obscurity proponents) quaked in fear at the disclosure of weapons for the hack community, other security researchers used it as a way to force security issues to get fixed, particularly the closed source ones. Eventually, after the Symantec acquisition of SecurityFocus, concern began that Bugtraq was not independent enough. An alternate list "Full-Disclosure" was created.

Ironically, one would think hackers would be in favor of disclosure; however, the Antisec movement began as a backlash to Bugtraq and the philosophy of full-disclosure. The original goals of the movement, which were published on their web site in 2001, appear below (Table 1).

The purpose of this movement is to encourage a new policy of anti-disclosure among the computer and network security communities. The goal is not to ultimately discourage the publication of all security-related news and developments, but rather, to stop the disclosure of all unknown or non-public exploits and vulnerabilities. In essence, this would put a stop to the publication of all private materials that could allow script kiddies from compromising systems via unknown methods.

The open-source movement has been an invaluable tool in the computer world, and we are all indebted to it. Open-source is a wonderful concept which should and will exist forever, as educational, scientific, and end-user software should be free and available to everybody.

Exploits, on the other hand, do not fall into this broad category. Just like munitions, which span from cryptographic algorithms to hand guns to missiles, and may not be spread without the control of export restrictions, exploits should not be released to a mass public of millions of Internet users. A digital holocaust occurs each time an exploit appears on Bugtraq, and kids across the world download it and target unprepared system administrators. Quite frankly, the integrity of systems world wide will be ensured to a much greater extent when exploits are kept private, and not published.

A common misconception is that if groups or individuals keep exploits and security secrets to themselves, they will become the dominators of the "illegal scene", as countless insecure systems will be solely at their mercy. This is far from the truth. Forums for information trade, such as Bugtraq, Packetstorm, www.hack.co.za, and vuln-dev have done much more to harm the underground and net than they have done to help them.

What casual browsers of these sites and mailing lists fail to realize is that some of the more prominent groups do not publish their findings immediately, but only as a last resort in the case that their code is leaked or has become obsolete. This is why production dates in header files often precede release dates by a matter of months or even years.

Another false conclusion by the same manner is that if these groups haven't released anything in a matter of months, it must be because they haven't found anything new. The regular reader must be made aware of these things.

We are not trying to discourage exploit development or source auditing. We are merely trying to stop the results of these efforts from seeing the light. Please join us if you would like to see a stop to the commercialization, media, and general abuse of infosec.

Thank you.

Table 1. Antisec first manifesto. Now offline. Retrieved from: http://web.archive.org/web/20010301215117/http://anti.security.is/

Here, the movement expresses the idea that security vulnerabilities are weapons, and disclose is tantamount to spreading munitions, particularly into the hands of script kiddies. An insightful observation from the open source debate is that closed source is really a mechanism of commerce rather than security [48]. Interpreting this manifesto in that light, it can be seen as economically motivated and protectionist: that true hackers want to keep vulnerabilities for themselves, for monetization, rather than make them available to those who haven't paid (money or sweat equity) for them.

In July 2009, the image hosting site, Imageshack, was used by the Antisec movement to spread a new version of their "manifesto." At this point, they have become much more explicit in their belief that the security industry itself was profiteering from the output of all of the freely disclosed security research and racketeering in spreading FUD (Fear, Uncertainty, and Doubt) about security (Figure 2).

Although the earlier manifesto is welcoming, entreating "please join us," the second one makes it clear "this isn't like before," and draws clear distinctions between "us" and "you." In addition to promoting rm -rf (delete everything) they show a much more militant side; quoting from their now off-line web site, http://www.anti-sec.com/ (from cache): "Fuck full-disclosure", "Fuck the security industry", "Hack everyone you can and then hack some more", "Own everyone", "Destroy everything", "Take down every public forum, group, or website that helps in promoting exploits and tools or have show-off sections."

This more militant iteration of the group is the one that began taking down web sites. One of the notable early examples being the Astalavista search portal. Over the past couple of years since that time, the aims of Antisec have diverged into more general anti-security hacktivism activities and formed the basis for the well-known hacktivist groups Anonymous and LulzSec [45]. At this point, the movement has lost touch with its roots in the polarizing debate over full disclosure. One of the key reasons is that vulnerabilties have been commercialized in the underground. Clearly, vulnerability disclosure lists are far less unlikely to be the release mechanism for zero-days.

One Microsoft researcher reviewed the historical full-disclosure lists, and concluded that attacks based on source code have been the exception rather than the rule, and that "the magnitude of this potential negative remains to be evaluated" [23]. In the end, full disclosure won, and the software industry has permanently changed to accommodate it. Now software projects often openly solicit attacks, during beta testing, and vendors actually fix security flaws when notified, before they are disclosed. This does not appear to make Antisec happier.

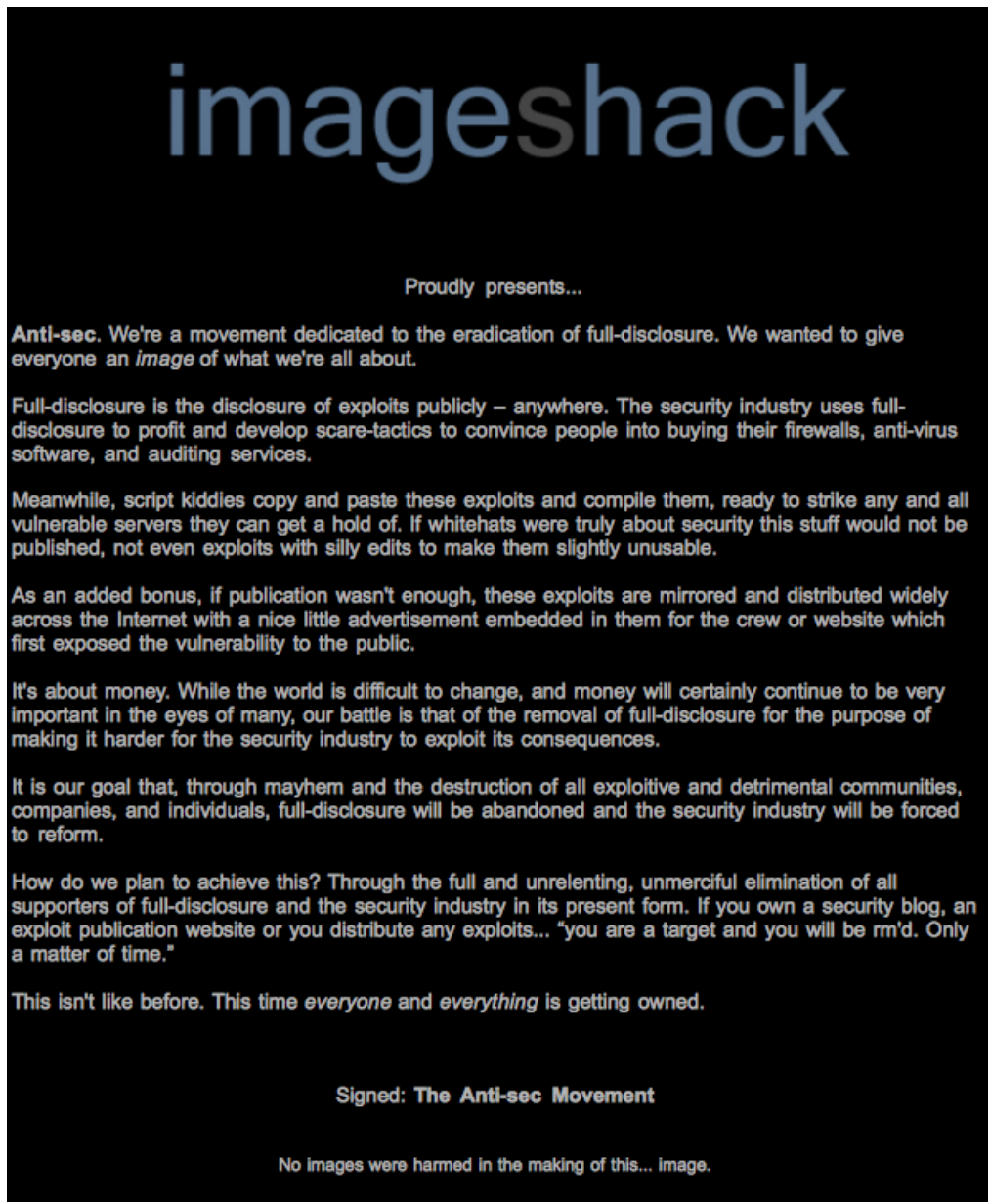

Figure 2. Retrieved from: http://mashable.com/2009/07/10/imageshack-hacked/

## 5. Obscurity Security: Faith?

An example of the classic view towards Security by obscurity is typified in [6]. Berghel believes that security in depth (SID) along with security through obscurity (STO) are both examples of "faith-based security": naively irrefutable yet with entirely no merit in practice. As he sees it, "The general premise of STO is that inviolability is a consequence of the enigmatic." Given such an argument, it is no wonder that he is able knock down this straw man.

However, no one is writing articles about the inviolability of security through obscurity, which they would certainly do if they so believed. It is easy to argue otherwise, drawing upon the old chestnut that there is no such thing as perfect security, only levels of risk. Both defense in depth as well as defense by secrecy (albeit temporary) provide a reduction in risk. Neither likely provides the quantity of risk reduction of a vetted cryptography algorithm, or a crowd-sourced (i.e. best practice) password policy. Yet it is a truism to say that until each layer of defense or secrecy is breached it is inviolate, by definition. That which changes in the quality of each defensive layer address is essentially the probability of that layer's breach. Those attacks that can't breach a specific layer are then filtered out; for them, that layer, and the goods inside, remains inviolate.

Clearly the disagreement we have here is a definition problem. Berghel lists three categories of security through obscurity, demonstrating the three by drawing upon mistakes from cryptography:

1. Failure to write secure code
2. Botched implementations
3. Inadequate education and training.

In examples where individuals in charge of security (like programmers) accidentally fall prey to these mistakes, it's hard to claim they willfully relied upon weak security, and name it "security by obscurity." Chances are they didn't know their systems were weak to being with. In fact, none of these three concepts are particularly relevant to those who willfully choose to employ obscurity measures.

One best practice in Windows is disabling and renaming the Administrator account. This would be an example of security through obscurity, yet it persists as universally recommended practice despite not being a sure preventative. Disabling and renaming the account just makes it harder for would-be attackers to get in. Here, harder provides value; it's a "stop stupid" deterrent. Similarly, the Windows security policy "Interactive: Do not display last user name" ensures that the previously used logon name is removed from the logon screen. Although a concerted attacker will not be deterred, why volunteer information? Taking these two implementation examples, can they be mapped to the list above? They cannot.

Failure to write secure code, is a tautological definition for those who choose obscure over secure code, and failed implementations of security which are trusted secure but flawed (botched), or attributable to a lack of education, are also not applicable to those who choose obscurity. These pitfalls are certainly to be avoided, and a thoughtfully chosen and implemented obscurity measure would be employed without triggering these mistakes. Clearly to be useful in practice, we need a definition for security by obscurity which is more empowering than depowering, and more appealing to the common sense and common practice proposition that reduced knowledge about a system will not help and could reasonably be expected to hinder an attacker.

## 6. Obscurity Nomenclature and Taxonomies

### 6.1. Definitions

So having defined ways that the security by obscurity term as been used destructively in the past, it is now illustrative to change focus to deriving constructive value from it. The Oxford English Dictionary defines the various forms of "obscurity" as making things unknown or undiscovered, that which conceals, is not clearly expressed, or not easily understood. Thinking about this definition for our purposes, it can be reduced to two attributes which might be applied for purposes of security, Firstly, we have the difficult to *find* (heretofore unknown, undiscovered, concealed), which is that case in which the object itself is obscured. And secondly, we have the difficult to *understand* (not clearly expressed, not easily understood), which is that case in which the meaning of the object is obscured. This second category also includes the example of lying about information, another form of obscuring the true meaning. If security by obscurity is to have value, we must find security value in making security-critical attributes hidden or unclear, not seeking to provide unattainable absolute secrecy, but increasing the difficulty in their finding or understanding.

### 6.2. Asymmetry

At the root of these themes is the idea of information *asymmetry*. When information is initially created it is an asymmetric state, known by the holder, controlled by the owner. Information is power. Information is a weapon. Maintaining this state of asymmetry keeps information, power, and weapons out of the hands of attackers. Ultimately this principle maps directly to one of the core values of the information security triad [44]: *confidentiality* (the others being integrity and availability).

One of the most basic advantages of secrecy lies in the time-to-market disadvantage it brings to security; this might be termed *time-to-attack*. Given a newly announced attack, from the abstract and theoretical, to the specific and leverageable, it might be fairly simple to search through large amounts of source code to find an application, and it is also a speedy and low-tech process to use tools from Google to grep to do so. On the other hand, finding an application for an exploit in a closed source product, such as Windows, or Acrobat, requires less common knowledge and skills, and more time, due to the complexity of the undertaking [4]. In this particular example from the closed-source vs. open source debate, the time-to-attack advantage to secrecy or obscurity is presumed to be offset by the higher rate of security problems, slower time-to-fix, and poorer coding practices associated with closed source software, and examples of these are readily found in the literature. This explanation from the one area of open/closed source is used to negate the general value of security by obscurity. However, this time-to-attack advantage does not exist only in a world of offsets, and it can be leveraged as a useful tool, if employed in way that does not come with the disadvantages attached to closed source. A careful consideration of the value that secrecy brings should reveal that it is not exactly in the nature of its secrecy, but rather in its obscurity.

Anderson [4] mathematically identifies an alternate aspect of asymmetry. He proposes that there are a large number of low-probability (read: obscure) bugs in software. The chance of an attacker finding any one successful attack is much higher than the defender having found that same security bug before the defender exploits it. This lack of symmetry is also a point against the conventional wisdom from the open source of the debate:

> Moreover, opening the source gives unfair advantage to the attacker. The attacker needs to find but one vulnerability to successfully attack the system. The defender needs to patch all vulnerabilities to protect himself completely. This is considered an uneven battle. [19]

In addition, there is one other major complication which is less often given attention in the literature, and further puts the defender at an enormous disadvantage: the asymmetry in the number of attackers (many), compared with the number of defenders (mostly, one) for a given instance of software. By definition, a SBO technique must by necessity leverage information asymmetry.

### 6.3. Conventional Wisdom

Highlighting the conventional wisdom can be illustrative of the properties to avoid as we look for effective security by obscurity.

Security by obscurity conventional wisdom:

- *SBO is misguided or ill conceived*

    Of course, we shall seek to avoid being wrong-headed.

- *Secrecy is a fallacy, and never lasts*

    The same can be said about security. Further, things that don't last still have value (time defeats all locks).

- *False confidence*

    We should seek to layer multiple security and obscurity measures and not require confidence in any single technique.

- *Lack of diagnostics and proof*

    This is the idea that security measures are tested in theory and/or practice, while obscurity measures are not. We will seek to show otherwise.

## 6.4. New Wisdom

We can hypothesize about some of the properties applicable in the realm of effective security by obscurity by going beyond the assumptions and considering which attributes are most likely to supply advantages and disadvantages.

Security by obscurity advantages:

- *Uniqueness*

    Customization defeats assumptions.

- *Hiding*

    Hiding knowledge makes obtaining knowledge about a system more difficult.

- *Deception*

    The truth is a set of one; lies are infinite. Attackers, like the rest of us, trust the information they have gathered firsthand.

- *Speed*

    True security, like cryptography, is slow. Obscurity used properly, tends to be easy, inexpensive, with no impact on performance.

- *Cost*

    True security costs more time and money to implement than obscurity.

Expanding on this last point, we must consider true security and security by obscurity to be different things, although they begin to converge (on one level) if both benefit from intentional design, SDLC integration, proven techniques and so forth. However, this is a good time to highlight that the have notably different *goals*. Security's goal is to be attack-proof, to defeat all attacks. Obscurity's goal, as we will see in more detail, attack-reducing, to defeat some attacks, which should be arbitrarily cheaper and more attainable.

Security by obscurity detriments:

- *Dependability*

  SBO is not (doesn't try to be) true security; it's a deterrent. Given a choice of one thing to implement, true security should be chosen. SBO is the better choice when compared to nothing, not when compared to true security.

- *Effort*

  Until automated, customization takes knowledge, will, and time. This is notably the case in initially attempting to use SBO, as many of the techniques are less well known.

- *Management*

  Anything unique, non-standard, customized, obscure, closed, etc..., is harder to manage and maintain than the well-known and open.

Considering all of the possible uses of security by obscurity, at this point a surfeit of terms and concepts have been identified. It seems counterproductive and premature to limit the characterization of techniques to a concise taxonomy, until more work analyzing these techniques has been performed.

# 7. Basic Obscurity Techniques

In the early days of internet warfare, a dominant method of target acquisition was via banner grabbing. SSH, Apache, Bind DNS, and many other servers were extended with run-time or build-time configuration options, allowing the misreporting of version information. This defeated many automated attacks, and even unskilled attackers (script kiddies, skids), who might then turn to an easier or more obviously vulnerable target. Although misdirection and false assurance are not true security, they can be true deterrents, and show one value of security by obscurity.

A list of simple security by obscurity techniques, many of which have been trivialized or obviated in practice might include such things as:

- Choose obscure software
- Lie about versions
- Obscure software configurations
- Obscure software locations
- Change defaults
- Change ports
- Change passwords
- Change banners
- Randomize attributes

Looking at the last item, an example of the harm that can be caused by static, predictable attributes is the Bloomberg incident, in November 2010, in which Bloomberg News obtained and released unpublished earnings of NetApp and Disney to its subscribers hours before official data releases. Although the pages were hidden, Bloomberg was able to trivially predict the resource locations [43]. Although real security would have been best, file name obscurity and/or randomization could have easily defeated (required brute force guessing) of this predictive attack on static URLs.

In response to the widespread adoption of hiding and evasion for web servers using the techniques above, among others, attackers have responded with HTTP fingerprinting technologies. One example is the httprint tool [38]. This method was specifically designed to defeat the obscurity methods of hiding information about web servers, relying on none of the customizable web server features, such as banners, strings, and other obfuscation. It relies on server and page signatures, attribute matching, statistical analysis, and fuzzy logic to identify web server versions.

In response to that, the industry has responded in this case by making layered security by obscurity a commercial enterprise. ServerMask is a relevant example in this case. They advertise themselves as "Anti-Reconnaissance for IIS Servers" focusing on defeating HTTP fingerprinting technologies. A list of features from their web site is included here (Table 2), and it is easy to see the preponderance of obscurity techniques being leveraged, simultaneously, by this software.

---

✓ **Stop Information Leakage: Web Server Anonymization**
- Obscure Headers, Cookies, & Error Messages

✓ **Remove Unnecessary HTTP Header & Response Data;**
- Broadcasting your Web server's identity allows intruders to complete their first task -- fingerprinting your technology. ServerMask removes unnecessary HTTP header and response data and camouflages your server by providing false signatures.

✓ **Eliminate File Extensions**
- File extensions like .asp or .aspx are clear indicators that a site is running on a Microsoft server. ServerMask eliminates the need to serve file extensions.

✓ **Modify Cookie Values**
- The ASP session ID cookie, used by the Session object to maintain client state, is a dead giveaway to the type of server you are running. ServerMask can modify your cookie values so that they are generic in nature and non identifiable.

✓ **Custom Error Pages**
- Default messages, pages and scripts of all kinds often contain clues to server identity. ServerMask custom error pages mask that information for better security.

✓ **Anti-Reconnaissance**
- Information masking encourages misguided exploits, snaring attackers with your firewalls and Intrusion Detection System. ServerMask augments these defenses to build more secure networks and return better results on security audits.

Table 2. ServerMask. http://www.port80software.com/products/servermask

---

Even in this basic area of security by obscurity (as in many others), despite the theoretical weakness of so many of these techniques individually, there is an arms race underway. We will consider more advanced obscurity techniques, in theory and practice, in later sections of the paper.

## 8. Provable Obscurity

One of the difficulties with using security by obscurity is the measurement of its effectiveness. A key problem is that attributes like uniqueness, customization, and deception have a tendency to make apples-to-apples comparisons impossible, and the controlling of variables necessary for bell-curve-based assessments infeasible. Yet only by quantifying the impact on SBO on vulnerability reduction can we answer questions about its efficacy.

### 8.1. Obscurity Probability Analysis

Fortunately, an individual implementation of, or an organization's customized approach to employing SBO lends itself to risk analysis. Risk analysis is always a case-by-case endeavor clearly unique to each asset being analyzed. Part of risk management involves taking calculated risks, and calculating risks entails a probabilistic approach to solving security issues. Simple probabilities can be leveraged to answer the question of the value of secrecy in this fashion. This technique stems from the very simple mathematics of risk analysis, which relies on estimated probabilities. For example, the factor of percent Annualized Rate of Occurrence (ARO), Asset Value (AV), and percent Exposure Factor (EF), create the Annual Loss Expectancy (ALE). Like so:

$$ALE = ARO \times (\text{AV} \times \text{EF})$$

Let's suppose that we take an intentional SBO approach to defending a specialized corporate web application. As this is a speculative exercise, we will not stop to document the assumptions which follow.

To begin, we choose to run our web server on the obscure NextStep OS (last released circa 1996), which we harden to have no externally accessible services other than this one. We decide to use thttpd, tiny httpd, a very lightweight open source web server, since this allows us to easily recompile it to support My Security Protocol, an SSL-like clone, using our own private encryption algorithm. We then compile it with a meaningless custom banner, and we set it to listen on port 18888. Finally, our application stores data locally, using Berkeley DB (BDB). Together each of these choices offers advantages and disadvantages for both usability and security, but we will focus on the latter, assuming we have complete control over our users' environment. When implementing specific choices like these it is important to also consider the wide reaching security ramifications they bring for the entire system; however we're focused here on the security of our web application, assuming all other avenues are either secure or irrelevant.

Imagine now that our web app faces a threat landscape consisting of 100,000 discrete attack sources, consisting equally of standalone *worms*, *bots* (botnets), *skids* (script kiddies), and *hackers* (humans). Let's say that, all other things being equal, our obfuscation of the banner alone completely deters some of these attacks, which merely check banners before launching; we lose from our threatscape (T), 5% of worms (.2w), 5% of bots (.2b), 5% of skids (.2s), and 1% (.01h) of hackers. Next, our choice of port 18888 filters out an additional group, expressed as [ .1w, .1b, .1s, .01h]. Our use of thttpd filters out [ .15w, .15b, .1s, .02h]. At this point, our selection of NextStep filters out an additional 10% of worms (.1w), 10% of bots (.1b), 10% of skids (.1s), and 20% (.2h) among the remaining hackers. Our choice of a proprietary protocol we've termed MSP is a deterrence for [.2w, .2b, .2s, .2h]. Finally, the use of BDB as a local database instead of MySQL or similar networked database, eliminates more attack methods, [.05w, .05b, .1s, .1h]. These quantities are additive, and would be trivially calculated like this:

$$T = 100,000 - (25,000 \times (.2 + .1 + .15 + .1 + .2 + .05)) - (25,000 \times (.2 + .1 + .15 + .1 + .2 + .05)) - ((25,000 \times (.2 + .1 + .1 + .1 + .2 + .1)) - ((25,000 \times (.01 + .01 + .02 + .2 + .2 + .1))$$

This can be reduced to:

$$T = 100,000 - 20,000(w) - 20,000(b) - 20,000(s) - 13,500(h) = 26,500.$$

As a result of our trivial security-obscurity changes, we've reduced the scope of the threat down to 26.5% of the initial attack set! Assuming, for purposes of illustration, that each attack source also takes time to find and time to compromise our system, let's say our average time to compromise given 100,000 attack sources, given our network location, speed, and external controls, began at 24 hours. Dropping the size of the threat group results in increasing the time-to-failure of this application to over 90 hours. This may not seem like much, but if we only plan to run the application for three days, it may indeed be good-enough security. Looked at from a risk management perspective, we might feed this reduction into our calculated estimate of the ARO and determine that the associated reduction in Annual Loss Expectancy, justifies our obscurity expenditure for these measures, and is more cost efficient than other approaches.

These cases admittedly have a degree of fantasy involved, but this method can be used in real cases to show real value for obscurity measures. All that's lacking is the true case-by-case measure of risk reduction that each method offers. Viewing these methods as valueless is what has prevented the formal study of the value of them, and doomed them to actually be valueless for most organizations, as they typically cannot be currently quantified for a risk analysis.

This filtering method, eliminating attackers is only one way to measure the value of SBO, one which makes the simplistic assumption that attackers will either a-priori know how to overcome an obstacle, or not, and be filtered out of attack group, at least for the current round of measurement. This tends to be the case for automated attack software, which would tend to match this black-and-white assumption of go-no go programmatic logic. Presumably attack software lacks attachment to neural networks and is incapable of learning to overcome a concrete obstacle!

However, this brings up the point that some attackers, some of the humans, will be capable of learning, and experience a delayed success in overcoming an attack obstacle. In these cases, SBO would serve to increase the work effort of attackers, rather than filter out the attack. Fortunately, there has been some treatment in the literature in analyzing the work effort effect on finding security flaws, in the area of software reliability engineering.

### 8.2. Security Failure Rate Analysis

One method of investigating the security failure rate of software was demonstrated by Anderson, who showed that the identification of software security bugs can be mathematically explained by extension from theoretical software reliability growth models [4]. Although that investigation focused on MTBF and the eradication of bugs in software by friendly agents, this same mathematical model can be applied to malicious actors -- also seeking to find bugs but for somewhat different purposes.

Through a polynomial distribution reduction, Anderson found that, given a consequence that makes the job of finding flaws on average $\lambda$ times harder, the probability of a security failure on the next test is reduced to $1/\lambda$. Harder, here, is defined as the average time spent finding a bug (security flaw). This method could therefore be used to mathematically describe the value of security by obscurity methods (used appropriately), in increasing the time required to find security vulnerability. Given the validity of these models to date, the following principles can be extrapolated to hold true, and presented as real benefits:

1. An obstacle which adds work, delays (increases) the time to security failure of software, decreasing the likelihood of failure over a given time.

2. So long as they are independent, obstacles which add work increase MTBF linearly (multiplicatively).

3. Increasing MTBF for software security adds value.

We now have identified two ways in which SBO benefits security. The first way was by reducing the size of the overall threat, in filtering out some attackers, specifically, some methods and sources. This property can be recast as a firewall (of sorts) for attackers. Just as a firewall that filters out packets on, say, ports 161-162, completely eliminate a certain class of attacks (SNMP attacks on standard ports), so does a SBO technique like port remapping or obfuscation completely eliminate attacker methods incapable of changing ports, or de-obfuscating code.

The second benefit is in increasing the time to failure (MTBF) of software security. MTBF has a long history of being an extremely useful security property. The most compelling example is in the area of cracking. In password encryption, quite often the entire security value of the encryption lies not in the security of the algorithm and implementation at all, but rather in the MTBF due to brute force attacking. An organization selects a set of parameters for their password security, such as length, complexity, and timers, which equate to an average expectancy of time before a given password is cracked on average. For example, it would take 5 days for an attacker to crack a 10-character password, drawn from lowercase, uppercase and numbers, using 100 GPUs [17].

Now, an organization who wishes to choose secure password settings might choose to mandate extra complexity for their employees' passwords. With the same assumptions, requiring a symbol in the set increases the work level such that the same attacker now requires 280 days to crack the password. Given forced password rotation every 90 days, this might be more than enough for the lifetime of the password and satisfy the MTBF needs of the organization. In a similar way, adding work level complexity to software – by way of obscurity techniques which impede security attacks – can be chosen such that the software is secure for its practical lifetime, which may be merely until the next recompilation, version, or patch.

## 9. So... Obscurity Security. Fact?

In one of the rare opinions in defense of security by obscurity, Stuttard proposes that "security measures thus dismissed are often worthwhile and can significantly mitigate the security threats facing the organization" [42]. The author lists four objectives which work in favor of security by obscurity:

1. Reducing the likelihood of an attack occurring.

2. Slowing down the progress of an attack.

3. Increasing the likelihood of an actual attack being detected.

4. Mitigating the impact of a successful attack.

The first of these arguments is similar to the properties I've presented. I argue that the reduction in likelihood of an attack occurring is due to the filtering property of obscurity measures, which reduce the attack base and decrease the size of threat. Likewise, increasing the difficulty of the attack, decreases the time-to-attack, and reduces the probability likelihood of a successful attack over a given time period. Stuttard explains that obscurity measures such as changing ports or renaming utilities can slow the progress of an attack; however, I hold it more likely that these measures will defeat automation, but fail to slow down manual attackers, who, to match the examples given, will already be engaged in scanning all ports, and not be relying in any great numbers on the presence of standard utilities. Since the time of writing, in 2005, malicious agents, from worms to black hats, have become accustomed to bootstrapping their own tools in place. There is also a lack of good examples for the properties of obscurity measures in increasing the likelihood of detection and assisting in mitigation of an attacks' impact. Stuttard's most valuable contribution is in presenting the flawed assumption behind the "real" security counter-argument: that security exists at all! The ideal behind it is simply not practically attainable, particularly in an age in which zero-days, social engineering, and a highly skilled attacker can defeat any security measure, given the time and willpower. What is practical instead is to put as many barriers as possible in the way of attack, and the intelligent use of obscurity measures to deter attack can be a valuable defensive layer alongside the more common measures like patching.

Stuttard attempts to debunk the standard "complacency" wisdom that an organization which thinks it has hidden its vulnerabilities behind obscurity measures might not be concerned about resolving them, by pointing out that many security by obscurity measures are technically sophisticated, like obfuscation or code morphology, and an organization which employs them likely has considered what weight to lend them amongst their other security measures. However, today's useful technically sophisticated skill is tomorrow's automated turn-key implementation. He also does not give ample weight to the many historical examples, such as the Clipper chip, or Netscape's failed POP password proprietary encryption [46], demonstrating that programmers might not be trustable after all to make smart choices, to seek out true security measures without external review, rather relying on proprietary secrecy to protect their data.

One promising research has analyzed the value of security by obscurity as it applies to game theory [30]. Pavlovic sees that two families of security practice, those to keep attackers out, and those to catch them when they are in, correspond to two families of strategies in certain games of incomplete information, and turn out to have opposite winning odds. When the defenders' goal is to keep the attackers out, it is observed that the attackers only need to find one attack vector to enter in static fortress mode, whereas the defenders must defend all attack vectors to prevent them.

When the battle is recast in a dynamic mode, then the defenders only need to find one marker to recognize and catch the attackers, whereas the attackers must cover all their markers. Therefore, by analyzing your enemy's behaviours and by obscuring your own, you can improve the odds of winning this game. Pavlovic proposes that security can be increased not only by analyzing attacker's type, but also by obscuring defender's type. Obscuring information is certainly a useful defensive technique as we will explore next.

## 10. Information Hiding

In examining the value of secrets, Diffie writes that "It isn't that secrets are never needed in security. It's that they are never desirable" [12]. Yet on closer inspection, he refines this statement with the assertion that, actually, they can't be relied upon for strong security, emphasis on strong. He posits that a secret which cannot be easily changed should be regarded as a vulnerability, due to the time and money necessary to design another system. It is evident that he is considering the case where this is a single secret, on which the security in use depends, and all value the secret provides is lost at its first disclosure. However, if you take a lesson from the DOD classification system, not all secrets are "Top secret," and classification is not the only way in which information is protected. The disclosure of "Confidential" information to an unapproved party is not a matter which results in the redesign of either the classification system or the roles and responsibilities assigned thereunder. Information is not even reclassified as "Unclassified" on its disclosure. The fact is, that making information public is not something that happens instantly. In applying Diffie's points to our SBO reasoning, where we are expecting multi- layered obscurity elements (secrets) and security elements together in use, the failure of one layer does not mean that all layers fail. In addition, despite the loss of a hidden secret to one attacker (like the administrator account name), the value of a security by obscurity measure will continue to perform in practice, against all of those attackers not in knowledge of, or reconfigured to overcome that obstacle. It is true that over long periods of time, secrets tend to get broken and disseminated. This is a good reason to consider obscurity measures as a better fit for short-term problems of protection.

And really, using concealment may be the only solution to some problems in security. For example, while encrypting a file provides true (data) security, hiding the very existence of a file requires concealment. Anane et al [1] present a novel method for file location concealment, through the fragmentation of files, their distribution across a network of nodes, and the removal of their meta-information. The expressed aims of concealment are in the service of enhancing secrecy, as well as enabling concealment as a service to users. Concealment can never be more than security by obscurity. These authors demonstrate an important value in pursuing SBO technologies: not only can they be leveraged to enhance the confidentiality principle of security, but users want them!

Security by obscurity by definition conceals information and systems by hiding them, however temporarily. Whereas cryptography can secure messages, it has little application for hiding their existence altogether [31]. Out of the field of Information Hiding comes techniques which, despite their reliance on "weak" obscurity, have proven useful application in a number of areas. For example, in multiple media and publishing industries, hiding techniques have come to be deployed to provide such capabilities as digital *watermarks* (hidden copyright messages), and *fingerprinting* (hidden serial numbers, for copyright control).

Information hiding has a long history of application in and a huge body of knowledge about its application in the theater of war. Keeping secrets, leveraging information imbalance, need to know, these are mature concepts in military circles. In the area of covert communications, hiding applications in the field of signals intelligence include techniques like spread spectrum radio to prevent transmitter location. In commodity communications systems, techniques like mobile subscriber identifiers provide location privacy. And an example from Internet communications systems includes proxies and anonymizers, which result in concealing a message sender's identity (such as in email).

For example, with the Crowds [33] system, anonymity through obscurity is provided by implementing the idea of "blending into a crowd." In hiding one's actions within the actions of many others, this model works like others in hiding information within other valid information. In this case, Crowds implements the attribute of privacy through the hiding principle, specifically lying, of security by obscurity. Crowds' "security" can be readily defeated by executable and other modular content which force a machine to open a direct connection or reveal its true identity via the Crowds proxy channel. However, unless specifically attacked and defeated in a transaction in this manner, Crowds' obscurity proves its value of anonymity by default in every other case. In the same way, other forms of SBO information hiding prove their value in each case where they were effective and have demonstrable superiority over a default state of information leakage.

Another well-known example is steganography (steganos Gk., covered), an important subdiscipline of information hiding. Steganography techniques conceal the existence of a message, most often by embedding hidden messages in other communications (the definition of message implies communication). Examples include classic real-world techniques like invisible ink on another message, as well as modern techniques, like hiding information in the least-significant bit of media files, a technique known to have been used by terrorist and intelligence organizations.

Whether covert communication channels or covert messages, these techniques benefit both malicious actors and good actors, who wish to hide the existence of their communications. Following Kerckhoffs, if a secure system is one where an adversary understands the system, but lacks the key, then information hiding supports that principle by hiding the fact that the system is in use, and such a key might be found, which is critical reconnaissance knowledge required to launch an attack.

A crucial component to understanding the value of security by obscurity is gauging its use in attackers' repertoires. Just because it is thought to hold little value for the security professional to leverage, does not mean that it is not used by attackers, and potentially potent against the unwary administrator. For example, recent research has studied data hiding inside Microsoft Office OOXML files [29], showing it to be an effective hiding scheme, and an important example of a category of awareness necessary for the security forensic investigator. Security by obscurity techniques need to be in the repertoire of every defender in order to offset the information asymmetry that would present if only attackers can leverage them. For example, an attacker might insert malicious code into an open source software project (or a malicious insider in a closed source), and use obscurity techniques to defeat detection via code review. Understanding this threat entails developing competency with the techniques, and how they can be used for good, as well as ill.

Any particular obscurity measure differs from a well-known measure in how widespread the knowledge is of its implementation. For example, steganography may be an effective method of hiding information, albeit a security by obscurity measure. Encoding information directly in binary in the least significant bit of each pixel of an image is not much of a deterrent if the overwhelming majority of practitioners choose to encode information this way. A successful attack on steganography consists in the detecting the existence of the communication and reversing the encoding. Research has shown that most if not all of the steganographic and covert systems in use today are trivial for a capable opponent to detect and remove [31]. The fact that these techniques remain obscure, helped by the diversity of their options and permutations, as well as the lack of capable opponents, is what enables those who employ them to stay one step ahead of their attackers.

Choosing a steganographic method which uses base64 for even and XOR for odd pixels, and alternates between final and penultimate least significant bit, while everyone continues to configure the previous way, is surely a stronger obscurity measure. Similarly, choosing a port like 8080 for an alternate (internal) web server port may be commonplace, and no longer obscure. Choosing a random port between 19000-49000 is much more likely to be obscure. Extending this strategy out, choosing port 29001 for all web servers in an organization provides less obscurity (particularly once one is located) than choosing a *different* port for each web server would provide. Of course, this could be a management burden in some organizations, but that could be balanced case-by-case. Extrapolating to an entire community, the authors of a particular piece of closed-source software might choose to obfuscate its object code, to aid in retaining a licensing or competitive advantage (as often happens). Choosing a different pattern of obfuscating mechanisms for each installation ensures that if one customer breaks the license, or a single competitor reverse engineers the code, making one instance public does not immediately weaken all installations of the software.

Essentially, we can see that there is another aspect to selecting a particular obscurity measure required to retain its property of obscurity and maximize its value: dynamism. A static choice offers least obscurity. This is one of the flaws in many security by obscurity measures in those coding practices which have been criticized in the past: that the using the same secret in each instance of code is least secure of all, and once it falls, all instances of it fail as well. Fortunately, coding lends itself to dynamic choices, and randomization measures would be the most reliable way to ensure a dynamic secret is chosen. This would not be difficult to implement. Following this practice through with our web server example, on installation or compilation the software might generate a random port number, a random banner, random fingerprint characteristics, and so on, offering them up to the user until all options (dynamic or static) have been selected. The individual would choose case-by-case whether a choice provides value in its obscurity, or provides attributes like interoperability and usability in its inobscurity.

## 11. Obfuscation and Metamorphism

"Total unintelligibility is the natural state of computer programs" [5], yet there are ways to make it even worse! Software obfuscation is a process which changes or hides the nature of software while retaining equivalent behavior. The field has many ties to cryptography, and notably, many algorithms make critical use of obfuscation techniques to supply data hiding characteristics. For example, the foundational concept of diffusion typically involve multiple rounds of activities (mixing, shifting) which provide obscurity rather than security – they are invertable – and are designed to add complexity. The concept of diffusion refers to distributing information widely to dissipate it and make it challenging to piece together [39], again merely obscuring it. Both of these security by obscurity concepts have been essential pieces in true cryptographic security and they both are techniques from the field of obfuscation.

Numerous ad-hoc heuristical techniques are used by practitioners every day to obfuscate their code, though few supply any provable notion of security [18]. Ideally, the goal of security obfuscation is to transform code so that the program runs the same as the original one, but that the original code (or its secrets) cannot be recovered. Perfect whole-code obfuscation would be more difficult than cryptography, because encrypting some data is less a challenge than encrypting all data and requiring it to still execute! After all, researchers have recently shown that perfect obfuscation is not provably achievable [5], [18], which distinguishes it from a truly secure (as in NP-complete) approach, like widely used cryptographic systems. Furthermore, reversing cryptography is typically a black-box endeavor, and an attacker does not get to inspect the decryption process. Reversing obfuscation on the other hand is clearly a white-box exercise, as the attacker can watch the code execute. Software obfuscation is therefore, arguably, a form of security by obscurity, as the secrecy of the obfuscation methods and parameters are all that protects the assets so obscured. Nonetheless obfuscation still has practical uses.

As typically employed, obfuscation has been used for such things as the protection of cryptographic algorithms, secret keys, proprietary code, software watermarks (often used in licensing), and for tamper prevention. This category of use, protecting secrets, is well known, and its value is proven [9]. One of the concerns with employing obfuscation for these purposes is that the tools will be obtained and used by attackers to reverse-engineer code protected by it. However even if fallible in a number of ways, obfuscation still provides a deterrence, particularly to the unprepared attacker, seeking to gain access to secret information. It takes preparation, time, and the will to identify the obfuscation method and reverse engineer it.

Obfuscation is a programming subdiscipline and includes techniques such as these:

- Rename all variables in code to arbitrary names
- Include code that never executes or that does nothing (confusion)
- Move code around, distribute (diffusion) or combine functions
- Use alternate character set encodings
- Encrypt program data parts

In software, obfuscation goals are actually counterintuitive, to make code hard to understand, and represent the opposite of good software engineering (e.g. spaghetti code). However, in security they make more sense. Obfuscation forces an attacker to invest more time and effort in analyzing code, essentially in defeating the obfuscation. The actual value is going to vary as a function of the attacker, and the level of obfuscation employed. Case in point: while there is plenty of empirical field evidence that reverse engineering works, given sufficient motivation, there is also sufficient evidence revealing it is a highly substantial deterrent. For example, while hacking out a buffer overflow in one object in Windows may be worth something, reverse engineering the entire Windows code set would be far more lucrative. Further were it merely a computationally difficult challenge, it would have fallen by the wayside in this area of cheap processing. The fact is that reversing code is an already extremely difficult endeavor, made even harder in the presence of metamorphism, obfuscation, and other information hiding techniques, which requires suitable tools, and the skill and experience to use them. This difficulty creates a level deterrence which is a significant security benefit, although neither reliable nor specifically quantifiable, in the general case.

Unfortunately, the computational challenge a technique like obfuscation poses to any given attacker is too localized to be reliably quantifiable, and the deterrent factor alone should not be relied upon for true security. On the plus side, obfuscation can be seen as another example of a sufficiently obscure or challenging technique which can serve as a filter: if only 50% of the set of attackers can reverse engineer the obfuscation, that does quantifiably reduce the threat environment.

Leveraging metamorphic obfuscation can dramatically limit the attack surface of a program. These techniques have already been proven quite effective for attackers themselves, in the development of anti-virus-evading malware. When metamorphic malware runs, it changes the opcode that's loaded into memory and then writes a new version of itself back to the infected host file. Without ever having the same sequence of native opcodes in memory, conventional signature-based detection is defeated, [28]. The most dramatic case of obfuscation, for example, would be one in which every installation or compilation of software leaves behind a different obfuscated binary. Typically, in binary obfuscation, control flow, data flow, and call graphs are created, and optional changes are applied, such as inserting code, re-ordering, renaming, and re-looping, in multiple passes.

An area which could use more study for employing obfuscation is in the avoidance of programmatic security weaknesses. It is true that one way it can help is by making programs difficult to understand for the attacker. However, many other classes of common attacks, rely on a byte-specific implementation of an attack. Theoretically, with a different allocation and location of variables and functions on the stack in each case, the attack surface can be limited to one instance.

Users of obfuscation techniques can substantially delay the location of such weaknesses in their software, particularly via automated attack tools, perhaps sufficiently to keep attackers focused on other, lower hanging fruit. Using metamorphic software principles for security requires ensuring that unique instances are all functionally (semantically) the same but differ in internal structure. Although all instances might contain the same buffer overflow, any specific buffer overflow attack will only work against some instances. This helps with break once, break every (BOBE) resistance.

In one example of such research, it was shown that compilers that randomize their output can defeat attacks that exploit specific, predictable vulnerabilities in compiled programs. The authors propose several techniques for obfuscation, including reordering code, and memory and linking randomization. Then they implement a proof-of-concept stack-reordering obfuscation which randomizes the amount stack memory allocation, and show how it defeats buffer-overflow attacks: a buffer overflow in one binary is not likely located in the same place in any other [16].

This research is another example reflecting how obscurity can defeat automated attacks. It was also a bellwether of later diversity research in showing that replicas exhibiting independence could be generated by running an obfuscator multiple times with different secret keys.

## 12. Obscurity and Diversity

Although some work has been done in the field of determining binary equivalence [35], there are not yet guaranteed techniques for determining the semantic equivalence of metamorphic code. The focus of this work tends to be on combating malware metamorphology [7], but, here, the lack of reliability of the known techniques is a decided shortcoming. On the other hand, research on semantic *difference* of binaries has been shown to be effective in reverse engineering patches [14].

Even in the absence of formal proofs of equivalence, good-enough diversity can be a valuable technique, albeit security by obscurity. Probably the most well-known and widely used example of a diversity mechanism is address space layout randomization (ASLR). Typical ASLR implementations randomize the base address of an executable, including the stack, as well as the heap and shared memory. The idea is to provide some unpredictability for locations of objects in memory so that an attack which relies on a static address of an object will fail. Address space randomization has been implemented in most major operating systems and was introduced in Windows Vista and Mac OS X 10.5.

The Diablo diversifier goes beyond simple obfuscation creating self modifying code, and also employs control flow obfuscation, code generation, folding, and unfolding [2]. This research found that while obfuscation complicates the analysis phase, tamper resistance complicates the modification phase, and diversity complicates the automation phase of an attack.

One recent research effort has focused on the creation of diverse executables as a formalized security mechanism for securing services [34]. Semantics-preserving code transformations, which the authors title *proactive obfuscation*, were used to provide independence from attack for instances or replicas of server processes. The idea is that this increases the work an adversary must undertake to compromise a service which employs executable instance replication, and limits the number of compromised instances. The authors take strategic note that this measure, like other security by obscurity techniques, creates a benefit which erodes over time, as adversaries manually generate customized attacks. Therefore, they introduce the idea of *epochs* where new instances are regenerated at each planned interval, using rapid re-obfuscation, and state transference. In so doing, they create a moving surface, in an attempt to stay one step ahead of the attacker.

## 13. Security is a Moving Target.

The field of research into diversity as a defensive strategy can be seen as example of a class of defenses which have been more recently been termed Moving Target (MT) technologies. The root of this idea can be seen in age-old military theory; the value of creating a moving target is one of the US Armed Forces' nine principles of war, first introduced in 1921:

> [Manuever] achieves results that would otherwise be more costly. Effective maneuver keeps enemies off balance by making them confront new problems and new dangers. Forces gain and preserve freedom of action, reduce vulnerability, and exploit success through maneuver. [Army Field Manual 3-0, Operations, 4-43]

The foundational concept associated with Moving Target defense is to take a lesson from one of the classic asymmetries of information security defense, that the defender bears the brunt of the effort and expense in protecting a large set of attack points, while the attacker only has to attack one (the right one) [19]. An attacker further has the advantage in that investments in reconnaissance and development pay off, since knowledge about static targets can be gathered over time. Also, a successful attack strategy works in many places, and multiple attacks are easily coordinated. MT theory represents an effort to flip that on its head, in forcing the attacker to attack a moving target, a large set of targets which represent only one protection point for the defender. In a way it takes advantage of a reverse race condition in the attack strategy, in which the time between each attack stage, locating a target, identifying the vulnerability, and exploiting the vulnerability, creates a time-of check, time-of-use flaw (TOCTOU), [46], which can be leveraged by a defender by invalidating the data. In addition, such efforts increase uncertainty and costs (in time and resources) for the attacker, and apparent complexity and diversity of targets, increasing the range of defense strategies available to the defender, as well as increasing resilience of the target through redundancy, reconfigurations, and mobility.

The primary goal of MT approaches is to create a dynamic and uncertain attack surface area of the system under attack, again, underlining the value of the security by obscurity dynamism, and information hiding characteristics we've discussed previously. Although not a new idea, it is actually a novel area of research, largely into mechanisms of protection whereby a system under attack changes configuration or location over time to evade or delay compromise. MT technologies are enabled through recent technical developments, such as virtualization and workload mobility, ubiquitous network connectivity, and just-in-time compilers. Research areas in this field range from network layer techniques, like networking configuration randomization or masking, to application layer techniques, like rotating application software stacks, to programming techniques, like instruction set randomization. A compendium of recent MT research on these and other topics can be found in [20].

Two criticisms can be levied on Moving Target defenses which we might (admittedly, a bit tongue-in-cheek) call *smoke and mirrors* and the *lottery effect*. In the first case, it is a truism that while some aspects of an attack surface are moving around, many others are not; in most cases, software is attached to the same host, or the same network stack, or the same resolver library, and so on. Circumventing the moving parts, and attacking through another vulnerability, reduces the method to smoke and mirrors. In the second case, the lottery effect, is based on the idea that, statistically, the invulnerability of the prize is based on the small size of the participant group. Given that the randomization employed in MT is a knowably finite set (the number of obfuscations or combinations is quite far from infinite), certainly given a sufficiently large number of attackers trying random methods, one will get "lucky." (Actually, it's only lucky from the individual's standpoint; it's predictive.) These criticisms seem to be par for the course with security by obscurity measures, but a careful consideration will show that they can be levied at just about any technique. For example, even good cryptography certainly fails if the attacker gains a password (in the first case) through social engineering, or (in the second case) by fortuitous guess.

Rather, the questionable value proposition of an MT technology relies on whether it can supply sufficiently unpredictable protection characteristics, and/or change characteristics faster than defenders can defeat each instance. Consequently, automatic diversification techniques, automated management for configurations, as well as randomization and dynamization techniques, are areas where research and progress is required, if what is a security by obscurity theoretical technique is to develop into a practical one.

Subsequent to the 2009 U.S. White House Cyberspace Policy Review, R&D into MTD technologies has become part of the Department of Homeland Security's mandate. The field is moving beyond research, as DHS sponsorship for Federal Business Opportunities in the field have just opened up, as of November, 2011 [11]. A key question remaining to be answered in practice is whether these strategies can be effected while creating security systems that are still sufficiently manageable to the administrator (and transparent to the user) as to enable their eventual uptake in real world security defensive practice.

## 14. Conclusion

Software security is at the root of computer and network security. From coding flaws to design flaws to architectural flaws, security vulnerabilities in software result in software-induced risks which exist at the root of most classes of system vulnerabilities. Explicitly designing software for security is required for good software engineering practices. One of the biggest threats to software security comes via the automation which is enabled by the automatic and ubiquitous network connectivity of computers today. Launching automated network attacks is trivial, and automated exploitation of classes of software weaknesses becomes trivial. Automation relies on well-known information: ports, programs, code, methods, etc.... With ubiquitous presence comes ubiquitous presentation of weaknesses, which automated attacks can leverage. The enemy of ubiquity is uniqueness, which often brings with it the power to defeat automation. Uniqueness is just one useful area of security by obscurity.

Another one of the promises of the techniques discussed in this paper, like reconfiguration, lying, information hiding, diversity, obfuscation, and moving target lies in their ability to supply evasion from attack, due to mobility, or due to lack of knowledge about defeating them. According to David Lacey, Jericho Forum founder "Yes, there is [security in obscurity]. Not everything is known or knowable to an attacker. This uncertainty prevents and deters the vast majority of attacks" [25].

Security by obscurity works. After all, not only do we know this intuitively, but it is also easy to come up with common-sense examples. Child-proof pill bottles keep children out through the obscurity of their mechanism (knowledge of the technique), combined with a lack of skill to defeat it (strength to squeeze). This does not mean that they impede every child, and the fact that some children can open them is not a reason to give up the technique, or invest in something more costly and more reliable, like locks for our pills. The security and the deterrent they provide is valuable for however long it lasts, in the cases where it works, despite the fact that it will someday be defeated by every child. The time delay thus enabled affords time (perhaps as much as 10 years) to solve the real security problem (user education!). This is great example of the filtering effect of an obscurity measure, and although it only filters out attackers who are children, that's good-enough security right there.

Moving to a more practical example, biometric authentication is one of the more piquant examples of a practical security by obscurity technique. For example, from the standpoint of a remote attack, an individual's fingerprints are definitely obscure. However, they are not truly secure since they are public -- we do wave them around in public and leave them lying around everywhere! We rely on the low probability of attack, and the work factor required; however, given will and time, these secrets can be acquired. Again, a key consideration to evaluating the security of this technique would be understanding your attacker; the risk is much greater if they live in your own house. Like many security by obscurity techniques, once the secret is exposed (all ten fingers) all of the security provided is gone with respect to that attack source, but all other attackers (who lack the fingerprints) would still find the security impenetrable. This is another case which points out the value of combining multiple layers of obscurity techniques: here, including other measures such as temperature and blood flow [46].

Academic curricula typically expose students to the idea of security by obscurity as a cautionary tale, via what amounts to simple examples of failure. Coverage of SBO in the literature, tends to repeat verbatim the conventional wisdom, making it challenging to locate unbiased, open perspectives on the subject. This is unfortunate, as many software security engineering areas are directly impacted by SBO considerations. They include: security principles, security designs, security risk management, defensive measures, common implementation weaknesses, and content protection. As we will have seen, issues in cryptography, information warfare, game theory, and even cyber politics are directly impacted by this issue. In actually, rather than a cautionary tale, security by obscurity acceptance as a useful technique is apparent in a number of exiting research areas, notably information hiding, diversity, obfuscation, and moving target defense research.

For security practitioners, effective security by obscurity perhaps highlights what they already know, that there is added value in the reconfiguration for proprietariness of their security systems. It is a useful part of an effective defense in depth approach, in which a security system has multiple secret "fall backs" that are used to increase the attacker cost of breaches [3]. We should also consider the issue of attack by obscurity. Not least of the reasons to explore security by obscurity is in maintaining knowledge of active threats. Many of the techniques drawn from these fields are being actively used in malware development, particularly, but not limited to, information hiding, and metamorphic code. Other techniques may be leveraged by an attacker to increase confusion and diminish information on the part of the defender, like transcoding, unexpected input, and code manipulations. Understanding the uses of effective security by obscurity techniques for both offense and defense will benefit any comprehensive security practice.

---

**Acknowledgements**

I would like to thank Natalia Ivenskaya and George Finney for supporting me with the freedom to pursue this work.

The formatting of this paper is courtesy of the Elsevier Procedia Computer Science template for Microsoft Word [15].

# References

[1] Anane, R., Dhillon, S. and Bordbar, B. Stateless data concealment for distributed systems, Journal of Computer and System Sciences, Volume 74, Issue 2, March 2008, Pages 243-254

[2] Anckaert, B.: Diversity for Software Protection. PhD thesis, Ghent University, 2008.

[3] Anderson, R. Security Engineering: A Guide to Building Dependable Distributed Systems. John Wiley & Sons, Inc., New York, 2001.

[4] Anderson, R. Security in Open versus Closed Systems - The Dance of Boltzmann, Coase, and Moore. In Open Source Software: Economics, Law and Policy, Toulouse, France, 2002.

[5] Barak, B., Goldreich, O., Impagliazzo, R., Rudich S., Sahai, A., Vadhan, S.P., and Yang, K. On the (im)possibility of obfuscating programs. In Proceedings of the 21st Annual International Cryptology Conference on Advances in Cryptology, CRYPTO '01, pages 1–18, London, UK, 2001. Springer-Verlag.

[6] Berghel, H. April 30, 2008. Faith-Based Security. Communications of the ACM 51, no. 4.

[7] Briones, I., and Gomez, A. Graphs, entropy and grid computing: Automatic comparison of malware. In Proceedings of the 2004 Virus Bulletin Conference, 2004.

[8] Brown, K. Opening the open source debate. Technical report, Alexis de Tocqueville Institution, June 2002.

[9] Collberg C., and Thomborson, C. Watermarking, Tamper-Proofing, and Obfuscation - Tools for Software Protection. *IEEE Transactions on Software Engineering*, 28(8):735-746, August 2002.

[10] Cowan, C. Software security for open-source systems. IEEE J. Security & Privacy, 1(1):38–45, 2003.

[11] Department of Homeland Security, Science and Technology (S&T) Directorate. H-SB012.1-002: Moving Target Defense. 2011.

[12] Diffie, W. Perspective: Decrypting The Secret to Strong Security. CNET Perspectives. http://news.com.com/2010-1071-980462.html, Jan. 2003.

[13] Duartenn, J. Comments on "Opening the Open Source Debate". Technical report, Security Skill Center, June 2002.

[14] Dullein, T., Rolles, R. Graph-based comparison of executable objects. In Proceedings of the Symposium sur la Securite des Technologies de L'information et des communications, 2005.

[15] Elsevier. Procedia Computer Science template for Microsoft Word. Retrieved from: http://www.elsevier.com/inca/publications/misc/ProcediaComputerScience_template.doc, November 25th, 2011.

[16] Forrest, S., Somayaji, A., and Ackley, D. Building Diverse Computer Systems. In Proc. HotOS VI, IEEE CSPress, Los Alamitos, Calif., 1997, pp. 67–72.

[17] Fossen, J. 2009. How long to crack a password spreadsheet. SANS. Retrieved from: http://www.sans.org/windows-security/2009/06/12/how-long-to-crack-a-password-spreadsheet

[18] Goldwasser S. and Rothblum, G. On Best-Possible Obfuscation. In TCC 2007: 194-213.

[19] Hoepman, J. January 31, 2007. Increased security through open source. Communications of the ACM 50, no. 1.

[20] Jajodia, S., Ghosh, A.K., Swarup, V., Wang, C., and Wang, X.S. Moving Target Defense: Creating Asymmetric Uncertainty for Cyber Threats. Advances in Information Security, Volume 54, Springer-Verlag, New York, 2011.

[21] Kearns, D. July 29, 2002. Security by obscurity? Network world 19, no. 30.

[22] Kerckhoffs, A. La cryptographie militaire. Journal des sciences militaires, IX, 1883. pp. 5–38, Jan. 1883, and pp. 161–191, Feb. 1883.

[23] Lipner, S. Security and Source Code Access: Issues and Realities. In Proceedings of IEEE Symposium on Security and Privacy, Oakland, CA. pp.124-125, 2000.

[24] Mercuri, R., and Neumann, P. Inside Risks: Security by obscurity. Comm. ACM, 46(11):160, December 2003.

[25] Messmer, E. November 10, 2008. Defining security myths, truisms. Network world 25, no. 44.

[26] Miller, B., Koski, D., Lee, C. P., Maganty, V., Murthy, R., Natarajan, A., and Steidl, J. Fuzz Revisited: A Re-examination of the Reliability of Unix Utilities and Services. Tech. Report, Computer Science Dept., Univ. of Wisconsin, Madison, 1995.

[27] Neumann, P. Robust Nonproprietary Software. In Proceedings of IEEE Symposium on Security and Privacy, Oakland, CA, 2000. pp.122~123.

[28] O'Kane, P.; Sezer, S.; McLaughlin, K. "Obfuscation: The Hidden Malware," Security & Privacy, IEEE , vol.9, no.5, pp.41-47, Sept.-Oct. 2011.

[29] Park, B., Park, J. and Lee, S. Data concealment and detection in Microsoft Office 2007 files, Digital Investigation, Volume 5, Issues 3-4, March 2009, Pages 104-114.

[30] Pavlovic, D. Gaming security by obscurity. 2011. New Security Paradigms Workshop 2011.

[31] Petitcolas, F., Anderson, R., Kuhn, M. Information hiding-a survey. Proceedings of the IEEE, 87(7), 1999, pp 1062-1078.

[32] Raymond, E. The cathedral and the bazaar. Knowledge, Technology & Policy. Springer Netherlands, 0897-1986, 12:3, 1999-09-07, pp 23-49**.**

[33] Reiter, M. and Rubin, A. Crowds: anonymity for Web transactions, ACM Transactions on Information and System Security 1 (1) (1998) 66–92.

[34] Roeder, T. and Schneider, F. B. Proactive obfuscation. ACM Transactions on Computing Systems 28(2) 2010.

[35] Sabin, T. Comparing binaries with graph isomorphisms. Bindview, http://razor.bindview.com/publish/papers/comparing-binaries.html, April 2004.

[36] Schneider, F.B. Open source in security: visiting the bizarre. in IEEE Symposium on Security and Privacy, Oakland, CA, 2000.

[37] Sezer, E., Ning, P., Kil, C., and Xu, J. MemSherlock: An Automated Debugger for Memory Corruption Vulnerabilities. In the Proceedings of 14th ACM Conference on Computer and Communication Security, Nov 2007.

[38] Shah, S. An introduction to HTTP fingerprinting. http://net-square.com/httprint/httprint paper.html, 2004.

[39] Shannon, C. Communication Theory of Secrecy Systems. Bell System Technical Journal, vol. 28(4), page 656–715, 1949.

[40] Stallman, R. Free Software, Free Society. Boston: GNU Press, 2002.

[41] Stevenson, F. Cryptanalysis of contents scrambling system, 1999. Retrieved from: http://insecure.org/news/cryptanalysis_of_contents_scrambling_system.htm

[42] Stuttard, D. July 31, 2005. Security & obscurity. Network security 2005, no. 7.

[43] Sun, F., Xu, L., and Su, Z. Static Detection of Access Control Vulnerabilities in Web Applications. In Proceedings of USENIX Security 2011, San Francisco, CA, August 8-12, 2011.

[44] Tittel, E., Chapple, M., and Stewart, J. 2003. CISSP: Certified Information Systems Security Professional. Sybex.

[45] Vichot, R. Doing it for the lulz?: Online Communities of Practice and Offline Tactical Media". Master's Thesis, Georgia Institute of Technology, May, 2009.

[46] Viega J., and McGraw, G. Building Secure Software: How to Avoid Security Problems the Right Way. Addison-Wesley, 2001.

[47] Wheeler, D. Why Open Source Software / Free Software (OSS/FS)? Look at the Numbers! http://www.dwheeler.com/oss_fs_why.html, 2007, accessed 2011-12-7.

[48] Witten, B.; Landwehr, C.; Caloyannides, M. Does open source improve system security? Software, IEEE , vol.18, no.5, pp.57-61, Sep/Oct 2001.